\documentclass[journal, draft, onecolumn]{IEEEtran}

\usepackage{cite}      
\usepackage{graphicx}  
\usepackage{psfrag}    
\usepackage{subfigure} 
\usepackage{url}       
\usepackage{amsmath}   
\usepackage{latexsym}
\usepackage{array}
\usepackage{epic}
\usepackage{eepicemu}

\newtheorem{definition}{Definition}
\newtheorem{example}{Example}

\newtheorem{theorem}{Theorem}

\begin{document}
%
\title{Interval Neutrosophic Logic}
%
%
\author{Haibin~Wang,~\IEEEmembership{}
        Florentin~Smarandache,~\IEEEmembership{}
        Yanqing~Zhang,~\IEEEmembership{Member,~IEEE}
        and~Rajshekhar~Sunderraman~\IEEEmembership{}
\thanks{
}}

%


\maketitle

\begin{abstract}
In this paper, we present the interval neutrosophic logic which generalizes 
the  interval valued fuzzy logic, intuitionistic 
fuzzy logic and paraconsistent logics. These logics only consider truth-degree
or falsity-degree of a proposition. In interval neutrosophic logic, we also
consider indeterminacy-degree which can capture more information under
uncertain situation.  
We will give the formal definition
of interval neutrosophic propositional calculus and interval neutrosophic
predicate calculus. Then we give one application of interval neutrosophic 
logic--Interval Neutrosophic Logic System (Interval Neutrosophic Logic Controller) to do approximate reasoning. Interval Neutrosophic Logic System consists of
neutrosophication, neutrosophic inference, neutrosophic rule base, neutrosophic
type reduction and deneutrosophication. The interval neutrosophic logic and
interval neutrosophic logic system can be applied to many potential real applications where information is imprecise, uncertain, incomplete and inconsistent
such as Web intelligence, medical informatics, bioinformatics, decision making,
etc. 
\end{abstract}


%
\IEEEpeerreviewmaketitle

\section{Introduction}

The concept of fuzzy sets was introduced by Zadeh in 1965~\cite{ZAD65}. Since
then fuzzy sets and fuzzy logic have been applied in many real applications
to handle uncertainty. The traditional fuzzy set uses one real number 
$\mu_{A}(x) \in [0, 1]$ to represent the grade of membership of fuzzy set $A$
defined on universe $X$. The corresponding fuzzy logic associates each 
proposition
$p$ a real number $\mu(p) \in [0, 1]$ which represents the degree of truth.
Sometimes $\mu_{A}(x)$ itself is uncertain and hard to be defined by a crisp
value. So the concept of interval valued fuzzy sets was proposed~\cite{TUR86}
to capture the uncertainty of grade of membership. Interval valued fuzzy set
uses an interval value $[\mu_{A}^L(x),\mu_{A}^U(x)]$ with 
$0 \leq \mu_{A}^L(x) \leq \mu_{A}^U(x) \leq 1$ to represent the grade of 
membership of fuzzy set. The traditional fuzzy logic can be easily extended
to the interval valued fuzzy logic. There are other efforts to extend the
classical fuzzy sets and fuzzy logic such as type-2 fuzzy sets and type-2
fuzzy logic~\cite{KM98,LM00,MJ02}. The family of fuzzy sets and fuzzy logic
can only handle ``complete" information that is if grade of truth-membership
is $\mu_{A}(x)$ then grade of false-membership is $1 - \mu_{A}(x)$ by default.
In some applications such as expert system, decision making system and 
information fusion, the information is both uncertainty and incomplete. That
is beyond the scope of fuzzy sets and fuzzy logic. In 1986, Atanassov
introduced the intuitionistic fuzzy sets~\cite{ATA86} which is a generalization
of fuzzy sets and provably equivalent to interval valued fuzzy sets. The
intuitionistic fuzzy sets consider both truth-membership and false-membership.
The corresponding intuitionistic fuzzy logic~\cite{ATA88,ATA90,ATA98} 
associates each proposition $p$ with two real number $\mu(p)$-truth degree
and $\nu(p)$-falsity degree respectively, where $\mu(p), \nu(p) \in [0, 1], \mu(p) + \nu(p) \leq 1$. So intuitionistic fuzzy sets and intuitionistic fuzzy
logic could handle uncertain and incomplete information.
 
However, the inconsitent information exist in a lot of real situations such
as those mentioned above. It is obvious that intuitionistic fuzzy logic
could not reason with inconsistency. Generally, two basic approaches have
been followed in solving the inconsistency problem in knowledge bases: 
belief revision and paraconsistent logics. The goal of the first approach
is to make an inconsistent theory consistent, either by revising it or by
representing it by a consistent semantics. On the other hand, the 
paraconsistent approach allows reasoning in the presence of inconsistency,
and contradictory information can be derived or introduced without 
trivialization~\cite{ACM02}. de Costa's ${\cal C}_w$ logic~\cite{COS77}
and Belnap's four-valued logic~\cite{BEL77} are two well-known paraconsistent
logics. 

Neutrosophy was introduced by Smarandache in 1995 and started from paradoxism
which was coined by him in 1980 and where he based the creation on utilization 
of contradictions, anththeses, oxymorons, paradoxes. Then it was a need for 
the characterization of paradoxes in logic, that's why he started the neutrosophy, which is the foundation of the neutrosophic logic and other neutrosophics 
where we can characterize a paradox. ``Neutrosophy is a branch
of philosophy which studies the origin, nature and scope of neutralities, as
well as their interactions with different ideational spectra"~\cite{SMA03}.
Neutrosophy includes neutrosophic probability, neutrosophic sets and 
neutrosophic logic. Neutrosophic sets (neutrosophic logic) is a powerful
general formal framework which generalize the concept of the classicl sets
(classical logic), fuzzy sets (fuzzy logic), intuitionistic fuzzy sets 
(intuitionistic fuzzy logic).
In neutrosophic set (neutrosophic logic), indeterminacy is 
quantified explicitly and truth-membership (truth-degree), 
indeterminacy-membership (indeterminacy-degree) and false-membership 
(falsity-degree) are independent. The independence assumption is very important
 in a lot of applications such as information fusion when we try to combine
the data from different sensors. It is different from intuitionistic fuzzy
sets (intuitionistic fuzzy logic) where indeterminacy membership is
$1 - \mu_{A}(x) - \nu_{A}(x)$ ($1 - \mu(p) - \nu(p)$) by default. 

The neutrosophic set generalizes the above mentioned sets from philosophical
point of view. From scientific or engineering point of view, the neutrosophic
set and set-theoretic operators need to be specified. Otherwise, it will be
difficult to apply in the real application. In~\cite{WPZR04} we discuss one 
kind of neutrosophic sets called interval neutrosophic sets
and define a type of set-theoretic operators but more ones can be defined. 
It is natural to define the interval
neutrosophic logic based on the concept of interval neutrosophic sets.
In this paper, we will give the formal definition of interval neutrosophic 
propositional calculus and interval neutrosophic predicate calculus.

The rest of paper is organized as follows. Section~\ref{INS} gives a brief
review of interval neutrosophic sets. Section~\ref{INPC1} gives the formal
definition of interval neutrosophic propositional calculus. 
Section~\ref{INPC2} gives the formal definition of interval neutrosophic
predicate calculus. Section~\ref{AAINL} provide one application example of
interval neutrosophic logic as the foundation for the design of interval
neutrosophic logic system.
In section~\ref{conclusion} we conclude the paper and discuss the possible
future research direction.

\section{Interval Neutrosophic Sets}
\label{INS}
This section gives a brief overview of concepts of interval 
neutrosophic set defined in~\cite{WPZR04}. Interval neutrosophic set is an
instance of neutrosophic set introduced in~\cite{SMA01} which can be used in
real scientific and engineering applications.

\begin{definition}[Interval Neutrosophic Set]
Let $X$ be a space of points (objects), with a generic element in $X$ denoted
by $x$. An interval neutrosophic set (INS) $A$ in $X$ is characterized by
truth-membership function $T_A$, indeterminacy-membership function $I_A$ and
false-membership function $F_A$. For each point $x$ in $X$, 
$T_A(x), I_A(x), F_A(x) \subseteq [0, 1]$.
\hfill{\space} $\Box$
\end{definition}
When $X$ is continuous, an INS $A$ can be written as
\[
   A = \int_{X} \langle T(x),I(x),F(x) \rangle / x, x \in X
\]
When $X$ is discrete, an INS $A$ can be written as
\[
   A = \sum_{i=1}^{n} \langle T(x_i),I(x_i),F(x_i) / x_i, x_i \in X
\]

\begin{example}
Consider parameters such as capability, trustworhiness, and price of semantic
Web services. These parameters are commonly used to define quality of service
of semantic Web services~\cite{WZR04}. Assume that $X = [x_1,x_2,x_3]$. $x_1$
is capability, $x_2$ is trustworthiness and $x_3$ is price. The values of
$x_1,x_2$ and $x_3$ are a subset of $[0,1]$. They are obtained from the 
questionaire of some domain experts, their option could be degree of good,
degree of indeterminacy and degree of poor. $A$ is an interval neutrosophic
set of $X$ defined by $A = \langle [0.2,0.4],[0.3,0.5],[0.3,0.5] \rangle / x_1
+ \langle [0.5,0.7],[0,0.2],[0.2],[0.3] \rangle / x_2 + \langle [0.6,0.8],[0.2,0.3],[0.2,0.3] \rangle / x_3$.
\hfill{\space} $\Box$
\end{example}
\begin{definition}
An interval neutrosophic set $A$ is empty if and only if its 
$\inf T_A(x) = \sup T_A(x) = 0, \inf I_A(x) = \sup I_A(x) = 1$ and
$\inf F_A(x) = \sup T_A(x) = 0$, for all $x$ in $X$.
\hfill{\space} $\Box$
\end{definition}

Let $A$ be an interval neutrosophic set on $X$, then 
$A(x) = \langle T_A(x), I_A(x), F_A(x) \rangle$. 
Let $\underline 0$ = $\langle 0, 0, 1 \rangle$ and 
$\underline 1$ = $\langle 1, 1, 0 \rangle$.
 
\begin{definition}
Let $A$ and $B$ be two interval neutrosophic sets defined on $X$.
$A(x) \leq B(x)$ if and only if 
\begin{eqnarray}
\inf T_A(x) \leq \inf T_B(x) &,& \sup T_A(x) \leq \sup T_B(x), \\
\inf I_A(x) \leq \inf I_B(x) &,& \sup I_A(x) \leq \sup I_B(x), \\
\inf F_A(x) \geq \inf F_B(x) &,& \sup F_A(x) \geq \sup F_B(x).
\end{eqnarray}
\hfill{\space} $\Box$
\end{definition}

\begin{definition}[Containment]
An interval neutrosophic set $A$ is contained in the other interval
neutrosophic set $B$, $A \subseteq B$, if and only if
$A(x) \leq B(x)$,
for all $x$ in $X$.
\hfill{\space}  $\Box$
\end{definition}

\begin{definition}
Two interval neutrosophic sets $A$ and $B$ are equal, written as $A = B$,
if and only if $A \subseteq B$ and $B \subseteq A$.
\hfill{\space} $\Box$
\end{definition}

Let $N = \langle [0,1] \times [0,1], [0,1] \times [0,1], [0,1] \times [0,1] \rangle$.

\begin{definition}[Complement]
Let $C_N$ denote a neutrosophic
\emph{complement} of $A$. Then $C_N$ is a function
\[
  C_N : N \rightarrow N 
\]
and $C_N$ must satisfing at least the following two axiomatic requirements:
\begin{enumerate}
\item $C_N(\underline 0)$ = $\underline 1$ and $C_N(\underline 1)$ = $\underline 0$ (boundary conditions).
\item Let $A$ and $B$ be two interval neutrosophic sets defined on $X$,  if
$A(x) \leq B(x)$, then $C_N(A(x)) \geq C_N(B(x)$, for all $x$ in $X$. (monotonicity).
\item Let $A$ be interval neutrosophic set defined on $X$, then
$C_N(C_N(A(x))) = A(x)$, for all $x$ in $X$. (involutivity).
\end{enumerate}
\hfill{\space} $\Box$
\end{definition}
 
There are many functions which satisfy the requirement to be the complement 
operator
of interval neutrosophic sets. Here we give one example.
\begin{definition}[Complement $C_{N_1}$]
The complement of an interval neutrosophic set $A$ is denoted by $\bar{A}$
and is defined by
\begin{eqnarray}
T_{\bar A}(x) &=& F_A(x), \\
\inf I_{\bar A}(x) &=& 1 - \sup I_{A}(x), \\
\sup I_{\bar A}(x) &=& 1 - \inf I_{A}(x), \\
F_{\bar A}(x) &=& T_A(x),
\end{eqnarray}
for all $x$ in $X$.
\hfill{\space}  $\Box$
\end{definition}

\begin{definition}[$N$-norm]
Let $U_N$ denote a neutrosophic
\emph{intersection} of two interval neutrosophic sets $A$ and $B$. 
Then $I_N$ is a function
\[
  I_N : N \times N \rightarrow N
\]
and $I_N$ must satisfing at least the following two axiomatic requirements:
\begin{enumerate}
\item $I_N(A(x), \underline 1) = A(x)$ (boundary condition).
\item $B(x) \leq C(x)$ implies $I_N(A(x), B(x)) \leq I_N(A(x), C(x))$ (monotonicity).
\item $I_N(A(x), B(x)) = I_N(B(x), A(x))$ (commutativity).
\item $I_N(A(x), I_N(B(x), C(x))) = I_N(I_N(A(x), B(x)), C(x))$ (associativity).
\end{enumerate}
for all $x$ in $X$.
\hfill{\space} $\Box$
\end{definition}

Here we give one example of intersection of two interval neutrosophic sets
which satisfies above $N$-norm axiomatic requirements. Many other
definitions could be given which depend on the applications.

\begin{definition}[Intersection $I_{N_1}$]
The intersection of two interval neutrosophic sets $A$ and $B$ is an interval
neutrosophic set $C$, written as $C = A \cap B$, whose truth-membership,
indeterminacy-membership and false-membership are related to those of $A$
and $B$ by
\begin{eqnarray}
\inf T_C(x) &=& \min(\inf T_A(x),\inf T_B(x)), \\
\sup T_C(x) &=& \min(\sup T_A(x),\sup T_B(x)), \\
\inf I_C(x) &=& \min(\inf I_A(x),\inf I_B(x)), \\
\sup I_C(x) &=& \min(\sup I_A(x),\sup I_B(x)), \\
\inf F_C(x) &=& \max(\inf F_A(x),\inf F_B(x)), \\
\sup F_C(x) &=& \max(\sup F_A(x),\sup F_B(x)),
\end{eqnarray}
for all $x$ in $X$.
\hfill{\space}  $\Box$
\end{definition}

\begin{definition}[$N$-conorm]
Let $I_N$ Let denote a neutrosophic
\emph{union} of two interval neutrosophic sets $A$ and $B$.
Then $U_N$ is a function
\[
  U_N : N \times N \rightarrow N
\]
and $U_N$ must satisfing at least the following two axiomatic requirements:
\begin{enumerate}
\item $U_N(A(x), \underline 0) = A(x)$ (boundary condition).
\item $B(x) \leq C(x)$ implies $U_N(A(x), B(x)) \leq U_N(A(x), C(x))$ (monotonic
ity).
\item $U_N(A(x), B(x)) = U_N(B(x), A(x))$ (commutativity).
\item $U_N(A(x), U_N(B(x), C(x))) = U_N(U_N(A(x), B(x)), C(x))$ (associativity).
\end{enumerate}
for all $x$ in $X$.
\hfill{\space} $\Box$
\end{definition}

Here we give one example of union of two interval neutrosophic sets
which satisfies above $N$-norm axiomatic requirements. Many other
definitions could be given which depend on the applications.

\begin{definition}[Union $U_{N_1}$]
The intersection of two interval neutrosophic sets $A$ and $B$ is an interval
neutrosophic set $C$, written as $C = A \cap B$, whose truth-membership,
indeterminacy-membership and false-membership are related to those of $A$
and $B$ by
\begin{eqnarray}
\inf T_C(x) &=& \max(\inf T_A(x),\inf T_B(x)), \\
\sup T_C(x) &=& \max(\sup T_A(x),\sup T_B(x)), \\
\inf I_C(x) &=& \max(\inf I_A(x),\inf I_B(x)), \\
\sup I_C(x) &=& \max(\sup I_A(x),\sup I_B(x)), \\
\inf F_C(x) &=& \min(\inf F_A(x),\inf F_B(x)), \\
\sup F_C(x) &=& \min(\sup F_A(x),\sup F_B(x)),
\end{eqnarray}
for all $x$ in $X$.
\hfill{\space}  $\Box$
\end{definition}

\begin{theorem}
\label{theorem1}
Let $P$ be the power set of all interval neutrosophic sets defined in
the universe X. Then $\langle P;I_{N_1}, U_{N_1} \rangle$ is a
distributive lattice. 
\end{theorem}
\begin{proof}
Let $A, B, C$ be the arbitrary interval neutrosophic sets defined on $X$.
It is easy to verify that $A \cap A = A, A \cup A = A$ (idempotency), 
$A \cap B = B \cap A, A \cup B = B \cup A$ (commutativity), $(A \cap B) \cap C = A \cap (B \cap C), (A \cup B) \cup C = A \cup (B \cup C)$ (associativity), and
$A \cap (B \cup C) = (A \cap B) \cup (A \cap C), A \cup (B \cap C) = (A \cup B) \cap (A \cup C)$ (distributivity). 
\end{proof}

\begin{definition}[Interval neutrosophic relation]
Let $X$ and $Y$ be two non-empty crisp sets. An interval neutrosophic relation
$R(X, Y)$ is a subset of product space $X \times Y$, and is characterized by
the truth membership function $T_{R}(x,y)$, indeterminacy membership function
$I_{R}(x,y)$ and falsity membership function $F_{R}(x,y)$, 
where $x \in X$ and $y \in Y$ and $T_{R}(x,y), I_{R}(x,y), F_{R}(x,y) \subseteq [0,1]$.
\end{definition}

\begin{definition}[Interval Neutrosophic Composition Functions]
The membership functions
for the composition of interval neutrosophic relation
$R(X,Y)$ and $S(Y,Z)$ is given by the \emph{interval neutrosophic sup-star 
composition} of $R$ and $S$ 
\begin{eqnarray}
T_{R \circ S} (x,z) &=& \sup_{y \in Y} \min(T_{R}(x,y), T_{S}(y,z)), \\
I_{R \circ S} (x,z) &=& \sup_{y \in Y} \min(I_{R}(x,y), I_{S}(y,z)), \\
F_{R \circ S} (x,z) &=& \inf_{y \in Y} \max(F_{R}(x,y), F_{S}(y,z)).
\end{eqnarray}
\end{definition}

If $R$ is an interval neutrosophic set rather than an interval neutrosophic
relation, then $Y = X$ and  \\
$\sup_{y \in Y} \min(T_{R}(x,y), T_{S}(y,z))$ becomes
$\sup_{y \in Y} \min(T_{R}(x), T_{S}(y,z))$, which is only a function of the output
variable $z$. 
It is similar for $\inf_{y \in Y} \max(I_{R}(x,y), I_{S}(y,z))$ and $\inf_{y \in Y} \max(F_{R}(x,y), F_{S}(y,z))$. Hence, the notation of 
$T_{R \circ S} (x,z)$ can be simplified to
$T_{R \circ S} (z)$, so that in the case of $R$ being just an interval neutrosophic set,
\begin{eqnarray}
T_{R \circ S} (z) &=& \sup_{x \in X} \min(T_{R}(x), T_{S}(x,z)), \\
I_{R \circ S} (z) &=& \sup_{x \in X} \min(I_{R}(x), I_{S}(x,Z)), \\
F_{R \circ S} (z) &=& \inf_{x \in X} \max(F_{R}(x), F_{S}(x,z)).
\end{eqnarray}

\begin{definition}[Truth-favorite]
The truth-favorite of interval neutrosophic set $A$ is an interval neutrosophic set $B$, written as $B = \triangle A$, whose truth-membership and false-membership are related to those of $A$ by
\begin{eqnarray}
\inf T_B(x) &=& \min(\inf T_A(x)+\inf I_A(x),1), \\
\sup T_B(x) &=& \min(\sup T_A(x)+\sup I_A(x),1), \\
\inf I_B(x) &=& 0, \\
\sup I_B(x) &=& 0, \\
\inf F_B(x) &=& \inf F_A(x), \\
\sup F_B(x) &=& \sup F_A(x),
\end{eqnarray}
for all $x$ in $X$.
\hfill{\space} $\Box$
\end{definition}

\begin{definition}[False-favorite]
The truth-favorite of interval neutrosophic set $A$ is an interval neutrosophic
set $B$, written as $B = \nabla A$, whose truth-membership and false-membersh
ip are related to those of $A$ by
\begin{eqnarray}
\inf T_B(x) &=& \inf T_A(x), \\
\sup T_B(x) &=& \sup T_A(x), \\
\inf I_B(x) &=& 0, \\
\sup I_B(x) &=& 0, \\
\inf F_B(x) &=& \min(\inf F_A(x)+\inf I_A(x),1), \\
\sup F_B(x) &=& \min(\sup F_A(x)+\sup I_A(x),1),
\end{eqnarray}
for all $x$ in $X$.
\hfill{\space} $\Box$
\end{definition}

\section{Interval Neutrosophic Propositional Calculus}
\label{INPC1}
In this section, we shall introduce the elements of an interval neutrosohpic
propositional calculus, basing our constructions on the definition of the
interval neutrosophic sets, and using the notations from the theory of
classical propositional calculus~\cite{MEN87}.

\subsection{Syntax of Interval Neutrosophic Propositional Calculus}
\label{SINPC11}
Here we give the formalization of syntax of interval neutrosophic
propositional calculus.

\begin{definition}
An \emph {alphabet} of interval neutrosophic propositional calculus 
consists of three classes of symbols:
\begin{enumerate}
\item A set of \emph {interval neutrosophic propositional variables}, denoted 
by lower-case letters, sometimes indexed;
\item Five \emph {connectives} $\wedge, \vee, \neg, \rightarrow, \leftrightarrow$ which are called the conjunction, disjunction, negation, implication, and
biimplication symbols respectively;
\item The parentheses ( and ).
\end{enumerate}
\hfill{\space}  $\Box$
\end{definition}

The alphabet of interval neutrosophic propositional calculus gives rise to
combinations, obtained by assembling connectives and interval neutrosophic
propositional variables in strings. The purpose of the construction rules is
to allow the specification of distinguished combinations, called formulas.

\begin{definition}
The set of formulas (well-formed formulas) of interval neutrosophic propositional calculus is defined
as follows.
\begin{enumerate}
\item Every \emph {interval neutrosophic propositioanl variable} is a formula;
\item If $p$ is a formula, then so is $(\neg p)$;
\item If $p$ and $q$ are formulas, then so are
      \begin{enumerate}
        \item $(p \wedge q)$,
        \item $(p \vee q)$,
        \item $(p \rightarrow q)$, and
        \item $(p \leftrightarrow q)$.
      \end{enumerate}
\item No sequence of symbols is a formula which is not required to be by 1, 2,
      and 3. 
\end{enumerate}
\hfill{\space}   $\Box$
\end{definition}

To avoid having formulas cluttered with parentheses, we adopt the following
precedence hierachy, with the hightest precedence at the top:
\begin{center}
$\neg$, \\
$\wedge, \vee$, \\
$\rightarrow, \leftrightarrow$.
\end{center}

Here is an example of interval neutrosophic propositional calculus formula:
\[
 \neg p_1 \wedge p_2 \vee (p_1 \rightarrow p_3) \rightarrow p_2 \wedge \neg p_3
\]

\begin{definition}
The \emph{language of interval neutrosophic propositional calculus} given by
an alphabet consists of the set of all formulas constructed from the symbols
of the alphabet.
\hfill{\space}  $\Box$
\end{definition}

\subsection{Semantics of Interval Neutrosophic Propositional Calculus}
\label{SINPC12}
The study of interval neutrosophic propositional calculus comprises, 
among others, a \emph {syntax}, which
allows the distinction of well-formed formulas, and a \emph {semantics}, the
purpose of which is the assignment of a meaning to well-formed formulas.

To each interval neutrosophic proposition $p$, we associate it with an ordered
triple components $\langle t(p), i(p), f(p) \rangle$, 
where $t(p), i(p), f(p) \subseteq [0, 1]$. $t(p), i(p), f(p)$ is called
truth-degree, indeterminacy-degree and falsity-degree of $p$ respectively.
Let this assignment be provided by an \emph{interpretation function} or
\emph{interpretation} $INL$ 
defined over
a set of propositions $P$ in such a way that
\[
  INL(p) = \langle t(p), i(p), f(p) \rangle.
\]
Hence, the function $INL: P \rightarrow N \times N \times N$ gives the 
truth, indeterminacy and falsity degrees of all propositions in $P$.
We assume that the interpretation function $INL$ assigns to the logical truth
$T: INL(T) = \langle 1,1,0 \rangle$, and to 
$F: INL(F) = \langle 0,0,1 \rangle$.

An interpretation which make a formula true is a \emph{model} of this formula.

Let $i, l$ be the subinterval of $[0,1]$. Then 
$i + l = [\inf i + \inf l, \sup i + \sup l]$, 
$i - l = [\inf i - \sup l, \sup i - \inf l]$,
$\max(i, l) = [\max(\inf i, \inf l), \max(\sup i, \sup l)]$,
$\min(i, l) = [\min(\inf i, \inf l), \min(\sup i, \sup l)]$.

The semantics of five interval neutrosophic proporsitional connectives is given
in Table I. Note that $p \leftrightarrow q$ if and only if $p \rightarrow q$ and $q \rightarrow p$. \\
\begin{table*}[hbtf]
\begin{center}
\label{table1}
\caption{Semantics of Five Connectives in Interval Neutrosophic Propositional Logic}
\begin{tabular}{c|c}
\hline
Connectives & Semantics \\
\hline
\hline
$INL(\neg p)$ & $\langle f(p), 1 - i(p), t(p) \rangle$ \\
\hline 
$INL(p \wedge q)$ &  
$\langle \min(t(p),t(q)), \min(i(p),i(q)), \max(f(p),f(q)) \rangle$ \\
\hline
$INL(p \vee q)$ & 
$\langle \max(t(p),t(q)), \max(i(p),i(q)), \min(f(p),f(q)) \rangle$ \\
\hline
$INL(p \rightarrow q)$ &
$\langle \min(1, 1-t(p)+t(q)), \min(1, 1-i(p)+i(q)), \max(0, f(q)-f(p)) \rangle$ \\
\hline
$INL(p \leftrightarrow q)$ & 
$\langle \min(1-t(p)+t(q),1-t(q)+t(p)),\min(1-i(p)+i(q),1-i(q)+i(p)),\max(f(p)-f(q),f(q)-f(p)) \rangle$ \\
\hline
\end{tabular}
\end{center}
\end{table*}

\begin{example} 
$INL(p) = \langle 0.5, 0.4, 0.7 \rangle$ and $INL(q) = \langle 1, 0.7, 0.2 \rangle$. Then, $INL(\neg p) = \langle 0.7, 0.6, 0.5 \rangle$, $INL(p \wedge \neg p) = \langle 0.5, 0.4, 0.7 \rangle$, $INL(p \vee q) = \langle 1, 0.7, 0.2 \rangle$, $INL(p \rightarrow q) = \langle 1, 0, 0 \rangle$. 
\hfill{\space} $\Box$
\end{example}

A given well-formed interval neutrosophic propositioanl formula will be called
a tautology (valid) if $INL(A) = \langle 1, 1, 0 \rangle$, for all 
interpretation
functions $INL$. It will be called a contradiction (inconsistent) if 
$INL(A) = \langle 0, 0, 1 \rangle$, for all interpretation functions $INL$.

\begin{definition}
Two formulas $p$ and $q$ are said to be \emph {equivalent}, denoted $p = q$,
if and only if the $INL(p) = INL(q)$ for every interpretation function $INL$.
\hfill{\space} $\Box$
\end{definition}

\begin{theorem}
\label{theorem2}
Let $F$ be the set of formulas and $\wedge$ be the meet and $\vee$ the join, 
then $\langle F; \wedge, \vee \rangle$ is a distributive lattice.
\end{theorem}
\begin{proof}
It is analogous to the proof of Theorem~\ref{theorem1}.
\end{proof}

\begin{theorem}
\label{theorem3}
If $p$ and $p \rightarrow q$ are tautologies, then $q$ is also a tautology.
\end{theorem}
\begin{proof}
Since $p$ and $p \rightarrow q$ are tautologies then for every $INL$,
$INL(p) = INL(p \rightarrow q) = \langle 1,1,0 \rangle$, that is \\
$t(p) = i(p) = 1, f(p) = 0$, 
$t(p \rightarrow q) = \min(1, 1 - t(p) + t(q)) = 1$,
$i(p \rightarrow q) = \min(1, 1 - i(p) + i(p)) = 1$, 
$f(p \rightarrow q) = \max(0, f(q) - f(p)) = 0$. Hence, \\
t(q) = i(q) = 1, f(q) = 0. So $q$ is a tautology. 
\end{proof}

\subsection{Proof Theory of Interval Neutrosophic Propositional Calculus}
\label{PTINPC1}
Here we give the proof theory for interval neutrosophic propositional logic
to complement the semantics.

\begin{definition}
The interval neutrosophic propositional logic is defined by the following
axiom schema.
\begin{center}
$p \rightarrow (q \rightarrow p)$ \\
$p_1 \wedge \ldots \wedge p_m \rightarrow q_1 \vee \ldots q_n$ provided some $p_i$ is some $q_j$ \\ 
$p \rightarrow (q \rightarrow p \wedge q)$ \\
$(p \rightarrow r) \rightarrow ((q \rightarrow r) \rightarrow (p \vee q \rightarrow r))$ \\
$(p \vee q) \rightarrow r$ iff $p \rightarrow r$ and $q \rightarrow r$ \\
$p \rightarrow q$ iff $\neg q \rightarrow \neg p$ \\
$p \rightarrow q$ and $q \rightarrow r$ implies $p \rightarrow r$ \\
$p \rightarrow q$ iff $p \leftrightarrow (p \wedge q)$ iff $q \rightarrow (p \vee q)$
\end{center}
\hfill{\space} $\Box$
\end{definition}

The concept of (formal) deduction of a formula from a set of formulas, that is,
using the standard notation, $\Gamma \vdash p$, is defined as usual; 
in this case,
we say that $p$ is a syntactical consequence of the formulas in $T$.

\begin{theorem}
For interval neutrosophic propositional logic, we have
\begin{enumerate}
\item $\{p\} \vdash p$,
\item $\Gamma \vdash p$ entails $\Gamma \cup \Delta \vdash p$,
\item if $\Gamma \vdash p$ for any $p \in \Delta$ and $\Delta \vdash q$, then
$\Gamma \vdash q$. 
\end{enumerate}
\end{theorem}
\begin{proof}
It is immediate from the standard definition of the syntactical consequence $(\vdash)$.
\end{proof}

\begin{theorem}
In interval neutrosophic propositional logic, we have:
\begin{enumerate}
\item $\neg \neg p \leftrightarrow p$
\item $\neg (p \wedge q) \leftrightarrow \neg p \vee \neg q$
\item $\neg (p \vee q) \leftrightarrow \neg p \wedge \neg q$ 
\end{enumerate}
\end{theorem}
\begin{proof}
Proof is straight forward by following the semantics of interval neutrosophic
propositional logic.
\end{proof}

\begin{theorem}
In interval neutrosophic propositional logic, the following schemas do not hold:\begin{enumerate}
\item $p \vee \neg p$
\item $\neg (p \wedge \neg p)$
\item $p \wedge \neg p \rightarrow q$
\item $p \wedge \neg p \rightarrow \neg q$
\item $\{p, p \rightarrow q\} \vdash q$
\item $\{p \rightarrow q, \neg q \} \vdash \neg p$
\item $\{p \vee q, \neg q \} \vdash p$
\item $\neg p \vee q \leftrightarrow p \rightarrow q$
\end{enumerate}
\end{theorem}
\begin{proof}
Immediate from the semantics of interval neutrosophic propositional logic.
\end{proof}

\begin{example}
To illustrate the use of the interval neutrosophic propositional consequence
relation consider the following example. 
\[
   p \rightarrow (q \wedge r)
\]
\[
   r \rightarrow s
\]
\[
   q \rightarrow \neg s
\]
\[
   a
\]
From $p \rightarrow (q \wedge r)$, we get $p \rightarrow q$ and $p \rightarrow r$. From $p \rightarrow q$ and $q \rightarrow \neg s$, we get $p \rightarrow \neg s$. From $p \rightarrow r$ and $r \rightarrow s$, we get $p \rightarrow s$. 
Hence, $p$ is equivalent to $p \wedge s$ and $p \wedge \neg s$.
However, we cannot detach $s$ from $p$ nor $\neg s$ from $p$. This is in part
due to interval neutrosophic propositional logic incorporating neither modus
ponens nor and elimination.
\hfill{\space} $\Box$
\end{example}

\section{Interval Neutrosophic Predicate Calculus}
\label{INPC2}
In this section, we will exend our consideration to the full language of
first order interval neutrosophic predicate logic. 
First we give the
formalization of syntax of first order interval neutrosophic predicate logic as
in classical first-order predicate logic.

\subsection{Syntax of Interval Neutrosophic Predicate Calculus}
\label{SINPC21}

\begin{definition}
An \emph {alphabet} of first order interval neutrosophic predicate calculus 
consists of seven classes of symbols:
\begin{enumerate}
\item \emph{variables}, denoted by lower-case letters, sometimes indexed;
\item \emph{constants}, denoted by lower-case letters;
\item \emph{function symbols}, denoted by lower-case letters, sometimes indexed;
\item \emph{predicate symbols}, denoted by lower-case letters, sometimes indexed;
\item Five \emph{connectives} $\wedge, \vee, \neg, \rightarrow, \leftrightarrow$ which are called the conjunction, disjunction, negation, implication, and 
biimplication symbols respectively;
\item Two \emph{quantifiers}, the \emph{universal quantifier} $\forall$ (for all) and the \emph{existential quantifier} $\exists$ (there exists);
\item The parentheses ( and ). 
\end{enumerate}
\hfill{\space} $\Box$
\end{definition}

To avoid having formulas cluttered with brackets, we adopt the following precedence hierachy, with the highest precedence at the top:
\begin{center}
\[
  \neg, \forall, \exists
\]
\[
  \wedge, \vee
\]
\[
  \rightarrow, \leftrightarrow
\]
\end{center}

Next we turn to the definition of the first order interval neutrosophic 
language given by an alphabet.

\begin{definition}
A \emph{term} is defined as follows:
\begin{enumerate}
\item A variable is a term.
\item A constant is a term.
\item If $f$ is an $n$-ary function symbol and $t_1, \ldots, t_n$ are terms,
then $f(t_1, \ldots, f_n)$ is a term.
\end{enumerate}
\hfill{\space} $\Box$
\end{definition}

\begin{definition}
A \emph{(well-formed )formula} is defined inductively as follows:
\begin{enumerate}
\item If $p$ is an $n$-ary predicate symbol and $t_1, \ldots, t_n$ are terms,
then $p(t_1, \ldots, t_n)$ is a formula (called an \emph{atomic formula} or,
more simply, an \emph{atom}).
\item If $F$ and $G$ are formulas, then so are $(\neg F), (F \wedge G), (F \vee G), (F \rightarrow G)$ and $(F \leftrightarrow G)$.
\item If $F$ is a formula and $x$ is a variable, then $(\forall x F)$ and
$(\exists x F)$ are formulas.
\end{enumerate}
\hfill{\space} $\Box$
\end{definition}

\begin{definition}
The \emph{first order interval neutrosophic language} given by an alphabet
consists of the set of all formulas constructed from the symbols of the alphabet.
\hfill{\space} $\Box$
\end{definition}

\begin{example}
$\forall x \exists y (p(x,y) \rightarrow q(x)), \neg \exists x (p(x,a) \wedge q(x))$ are formulas.
\hfill{\space} $\Box$
\end{example} 

\begin{definition}
The \emph{scope} of $\forall x$ (resp. $\exists x$) in $\forall x F$ (resp. $\exists x F$) is $F$. A \emph{bound occurrence} of a variable in a formula is an
occurrence immediately following a quantifier or an occurrence within the 
scope of a quantifier, which has the same variable immediately after the 
quantifier. Any other occurrence of a variable is $\emph{free}$.
\hfill{\space} $\Box$
\end{definition}

\begin{example}
In the formula $\forall x p(x,y) \vee q(x)$, the first two occurrences of $x$
are bound, while the third occurrence is free, since the scope of $\forall x$
is $p(x,y)$ and $y$ is free.
\hfill{\space} $\Box$
\end{example}

\subsection{Semantics of Interval Neutrosophic Predicate Calculus}
\label{SINPC22}
In this section, we study the semantics of interval neutrosophic predicate
calculus, the purpose of which is the assignment of a meaning to well-formed
formulas.
In the interval neutrosophic propositional logic, an interpretation is an
assignment of truth values (ordered triple component) to propositions. 
In the first order interval neutrosophic predicate logic, since there are
variables involved, we have to do more than that. To define an interpretation
for a well-formed formula in this logic, we have to specify two things, the 
domain and an assignment to constants and predicate symbols occurring in the
formula. The following is the formal definition of an interpretation of a
formula in the first order interval neutrosophic logic.

\begin{definition}
An \emph{interpretation function (or interpretation)} of a formula $F$ in the
first order interval neutrosophic predicate logic consists of a nonempty domain
$D$, and an assignment of ``values" to each constant and predicate symbol
occurring in $F$ as follows:
\begin{enumerate}
\item To each constant, we assign an element in $D$.
\item To each $n$-ary function symbol, we assign a mapping from $D^n$ to $D$.
(Note that $D^n = \{(x_1, \ldots, x_n) | x_1 \in D, \ldots, x_n \in D \}$).
\item Predicate symbols get their meaning trough evaluation functions $E$ which 
assign to each variable $x$ an element $E(x) \in D$. 
To each $n$-ary predicate symbol $p$, there is a function 
$INP(p): D^n \rightarrow N \times N \times N$. So 
$INP(p(x_1, \ldots, x_n)) = INP(p)(E(x_1), \ldots, E(x_n))$. 
\end{enumerate}
\hfill{\space}  $\Box$
\end{definition}

That is, $INP(p)(a_1, \ldots, a_n) = \langle t(p(a_1, \ldots, a_n)), i(p(a_1, \ldots, a_n)), f(p(a_1, \ldots, a_n))$, \\ 
where $t(p(a_1, \ldots, a_n)), i(p(a_1, \ldots, a_n)), f(p(a_1, \ldots, a_n)) \subseteq [0,1]$. They are called truth-degree, indeterminacy-degree and falsity-degree of $p(a_1, \ldots, a_n)$ respectively. We assume that the interpretation function $INP$ assigns to the logical truth $T: INP(T) = \langle 1, 0, 0 \rangle$, and to $F: INP(F) = \langle 0, 1, 1 \rangle$. 

The semantics of five interval
neutrosophic predicate connectives and two quantifiers is given in Table II. For simplication of notation, we use $p$ to denote $p(a_1, \ldots, a_i)$.Note that $p \leftrightarrow q$ if and only if $p \rightarrow q$ and $q \rightarrow p$.

\begin{table*}[hbtf]
\begin{center}
\label{table2}
\caption{Semantics of Five Connectives and Two Quantifiers in Interval Neutrosophic Predicate Logic}
\begin{tabular}{c|c}
\hline
Connectives & Semantics \\
\hline
\hline
$INP(\neg p)$ & $\langle f(p), 1 - i(p), t(p) \rangle$ \\
\hline
$INP(p \wedge q)$ &
$\langle \min(t(p),t(q)), \min(i(p),i(q)), \max(f(p),f(q)) \rangle$ \\
\hline
$INP(p \vee q)$ &
$\langle \max(t(p),t(q)), \max(i(p),i(q)), \min(f(p),f(q)) \rangle$ \\
\hline
$INP(p \rightarrow q)$ &
$\langle \min(1, 1-t(p)+t(q)), \min(1,1-i(p)+i(q)), \max(0, f(q)-f(p)) \rangle$ \\
\hline
$INP(p \leftrightarrow q)$ &
$\langle \min(1-t(p)+t(q),1-t(q)+t(p)),\min(1-i(p)+i(q),1-i(q)+i(p)), \max(f(p)-f(q
),f(q)-f(p)) \rangle$ \\
\hline
$INP(\forall x F)$ & $\langle \min t(F(E(x))), \min i(F(E(x))), \max f(F(E(x))) \rangle$, $E(x) \in D$ \\
\hline 
$INP(\exists x F)$ & $\langle \max t(F(E(x))), \max i(F(E(x))), \min f(F(E(x))) \rangle$, $E(x) \in D$ \\
\hline 
\end{tabular}
\end{center}
\end{table*}

\begin{example}
Let $D = {1,2,3}$ and $p(1) = \langle 0.5,1,0.4 \rangle, p(2) = \langle 1,0.2,0 \rangle, p(3) = \langle 0.7,0.4,0.7 \rangle$. 
Then $INP(\forall x p(x)) = \langle 0.5,0.2,0.7 \rangle$, 
and $INP(\exists x p(x)) = \langle 1,1,0 \rangle$. 
\hfill{\space} $\Box$
\end{example}

\begin{definition}
A formula $F$ is \emph{consistent (satisfiable)} if and only if there exists
an interpretation $I$ such that $F$ is evaluated to $\langle 1,1,0 \rangle$ in 
$I$. if a formula $F$ is $T$ in an interpretation $I$, we say that $I$ is a 
\emph{model} of $F$ and $I$ \emph{satisfies} $F$.
\hfill{\space}  $\Box$
\end{definition}

\begin{definition}
A formula $F$ is \emph{inconsistent (unsatisfiable)} if and only if there 
exists no interpretation that satisfies $F$.
\hfill{\space} $\Box$
\end{definition}

\begin{definition}
A formula $F$ is \emph{valid} if and only if every interpretation of $F$ satisfies $F$.
\hfill{\space} $\Box$
\end{definition}

\begin{definition}
A formula $F$ is a \emph{logical consequence} of formulas $F_1, \ldots, F_n$ if
and only if for every interpretation $I$, if $F_1 \wedge \ldots \wedge F_n$ is
true in $I$, $F$ is also true in $I$.
\hfill{\space}  $\Box$
\end{definition}

\begin{example}
$(\forall x)(p(x) \rightarrow (\exists y)p(y)$ is valid, $(\forall x)p(x) \wedge (\exists y) \neg p(y)$ is consistent.
\hfill{\space}  $\Box$
\end{example}

\begin{theorem}
There is no inconsistent formula in first order interval neutrosophic predicate logic.
\end{theorem}
\begin{proof}
It is direct from the definition of semantics of interval neutrosophic predicate logic.
\end{proof}

Note that the first order interval neutrosophic predicate logic can be considered as an extension of the interval neutrosophic propositional logic. When a
formula in the first order logic contains no variables and quantifiers, it
can be treated just as a formula in the propositional logic.

\subsection{Proof Theory of Interval Neutrosophic Predicate Calculus}
\label{PTINPC2}
In this part, we give the proof theory for first order interval neutrosophic
predicate logic to complement the semantics part.

\begin{definition}
The first order interval neutrosophic predicate logic is defined by the following axiom schema.
\begin{center}
$(p \rightarrow q(x)) \rightarrow (p \rightarrow \forall x q(x))$ \\
$\forall x p(x) \rightarrow p(a)$ \\
$p(x) \rightarrow \exists x p(x)$ \\
$(p(x) \rightarrow q) \rightarrow (\exists x p(x) \rightarrow q)$ 
\end{center}
\hfill{\space} $\Box$
\end{definition}

\begin{theorem}
In first order interval neutrosophic predicate logic, we have:
\begin{enumerate}
\item $p(x) \vdash \forall x p(x)$
\item $p(a) \vdash \exists x p(x)$
\item $\forall x p(x) \vdash p(y)$
\item $\Gamma \cup \{p(x)\} \vdash q$, then $\Gamma \cup \{\exists x p(x)\} \vdash q$
\end{enumerate}
\end{theorem}
\begin{proof}
Directly from the definition of the semantics of first order interval 
neutrosophic predicate logic. 
\end{proof}

\begin{theorem}
In first order interval neutrosophic predicate logic, the following 
schemes are valid, where $r$ is a formula in which $x$ does not appear free:
\begin{enumerate}
\item $\forall x r \leftrightarrow r$
\item $\exists x r \leftrightarrow r$
\item $\forall x \forall y p(x,y) \leftrightarrow \forall y \forall x p(x,y)$
\item $\exists x \exists y p(x,y) \leftrightarrow \exists y \exists x p(x,y)$
\item $\forall x \forall y p(x,y) \rightarrow \forall x p(x,x)$
\item $\exists x p(x,x) \rightarrow \exists x \exists y p(x,y)$
\item $\forall x p(x) \rightarrow \exists x p(x)$
\item $\exists x \forall y p(x,y) \rightarrow \forall y \exists x p(x,y)$
\item $\forall x (p(x) \wedge q(x)) \leftrightarrow \forall x p(x) \wedge \forall x q(x)$
\item $\exists x (p(x) \vee q(x)) \leftrightarrow \exists x p(x) \vee \exists x q(x)$
\item $p \wedge \forall x q(x) \leftrightarrow \forall x (p \wedge q(x))$
\item $p \vee \forall x q(x) \leftrightarrow \forall x (p \vee q(x))$
\item $p \wedge \exists x q(x) \leftrightarrow \exists x (p \wedge q(x))$
\item $p \vee \exists x q(x) \leftrightarrow \exists x (p \vee q(x))$

\item $\forall x (p(x) \rightarrow q(x)) \rightarrow (\forall x p(x) \rightarrow \forall x q(x))$
\item $\forall x (p(x) \rightarrow q(x)) \rightarrow (\exists x p(x) \rightarrow \exists x q(x))$ 
\item $\exists x (p(x) \wedge q(x)) \rightarrow \exists x p(x) \wedge \exists x q(x)$
\item $\forall x p(x) \vee \forall x q(x) \rightarrow \forall x (p(x) \vee q(x))$
\item $\neg \exists x \neg p(x) \leftrightarrow \forall x p(x)$
\item $\neg \forall x \neg p(x) \leftrightarrow \exists p(x)$
\item $\neg \exists x p(x) \leftrightarrow \forall x \neg p(x)$
\item $\exists x \neg p(x) \leftrightarrow \neg \forall x p(x)$
\end{enumerate}
\end{theorem}
\begin{proof}
It is straightforward from the definition of the semantics and axiomatic 
schema of first order interval neutrosophic predicate logic.
\end{proof}

\section{An Application of Interval Neutrosophic Logics}
\label{AAINL}
In this section we provide one practical application of interval neutrosophic
logic -- Interval Neutrosophic Logic System (INLS). 
INLS could handle rule 
uncertainty as same as type-2 FLS~\cite{LM00}, besides, it could handle rule
inconsistency without the danger of trivilization. Like the classical FLS,
INLS is also characterized by IF--THEN rules. INLS consists of 
neutrosophication, neutrosophic inference, a neutrosophic rule base, 
neutrosophic type reduction and deneutrosophication. Given an input vector 
$x = (x_1, \ldots, x_n)$, where $x_1, \ldots, x_n$ can be crisp inputs or 
neutrosophic sets, the INLS
will generate a crisp output $y$. The general scheme of INLS
is shown in Fig. 1.

\begin{figure}[htbp]
\label{figure1}
\begin{center}

\setlength{\unitlength}{0.00083333in}
\begingroup\makeatletter\ifx\SetFigFont\undefined%
\gdef\SetFigFont#1#2#3#4#5{%
  \reset@font\fontsize{#1}{#2pt}%
  \fontfamily{#3}\fontseries{#4}\fontshape{#5}%
  \selectfont}%
\fi\endgroup%
{\renewcommand{\dashlinestretch}{30}
\begin{picture}(5648,2514)(0,-10)
\path(2112,1662)(2862,1662)(2862,1137)
	(2112,1137)(2112,1662)
\path(2112,837)(2862,837)(2862,312)
	(2112,312)(2112,837)
\path(2862,1362)(3237,1362)
\blacken\path(3117.000,1332.000)(3237.000,1362.000)(3117.000,1392.000)(3117.000,1332.000)
\path(2487,837)(2487,1137)
\blacken\path(2517.000,1017.000)(2487.000,1137.000)(2457.000,1017.000)(2517.000,1017.000)
\path(1737,1362)(2112,1362)
\blacken\path(1992.000,1332.000)(2112.000,1362.000)(1992.000,1392.000)(1992.000,1332.000)
\path(912,1662)(1737,1662)(1737,1137)
	(912,1137)(912,1662)
\path(762,2487)(1887,2487)(1887,12)
	(762,12)(762,2487)
\path(912,2187)(1737,2187)(1737,1812)
	(912,1812)(912,2187)
\path(1737,1962)(1887,1962)
\path(912,837)(1737,837)(1737,312)
	(912,312)(912,837)
\path(1737,612)(1887,612)
\path(762,612)(912,612)
\blacken\path(792.000,582.000)(912.000,612.000)(792.000,642.000)(792.000,582.000)
\path(12,1362)(912,1362)
\blacken\path(792.000,1332.000)(912.000,1362.000)(792.000,1392.000)(792.000,1332.000)
\path(762,2037)(912,2037)
\blacken\path(792.000,2007.000)(912.000,2037.000)(792.000,2067.000)(792.000,2007.000)
\path(3237,1662)(3837,1662)(3837,1137)
	(3237,1137)(3237,1662)
\path(3837,1362)(4212,1362)
\blacken\path(4092.000,1332.000)(4212.000,1362.000)(4092.000,1392.000)(4092.000,1332.000)
\path(4212,1662)(5037,1662)(5037,1137)
	(4212,1137)(4212,1662)
\path(5037,1362)(5562,1362)
\blacken\path(5442.000,1332.000)(5562.000,1362.000)(5442.000,1392.000)(5442.000,1332.000)
\put(2262,612){\makebox(0,0)[lb]{\smash{{{\SetFigFont{6}{7.2}{\rmdefault}{\mddefault}{\updefault}Neutrosophic}}}}}
\put(2262,462){\makebox(0,0)[lb]{\smash{{{\SetFigFont{6}{7.2}{\rmdefault}{\mddefault}{\updefault}Rule Base}}}}}
\put(1062,1512){\makebox(0,0)[lb]{\smash{{{\SetFigFont{6}{7.2}{\rmdefault}{\mddefault}{\updefault}Indeterminacy-}}}}}
\put(1062,1362){\makebox(0,0)[lb]{\smash{{{\SetFigFont{6}{7.2}{\rmdefault}{\mddefault}{\updefault}membership}}}}}
\put(1062,1212){\makebox(0,0)[lb]{\smash{{{\SetFigFont{6}{7.2}{\rmdefault}{\mddefault}{\updefault}Function}}}}}
\put(987,2337){\makebox(0,0)[lb]{\smash{{{\SetFigFont{6}{7.2}{\rmdefault}{\mddefault}{\updefault}Neutrosophication}}}}}
\put(987,2037){\makebox(0,0)[lb]{\smash{{{\SetFigFont{6}{7.2}{\rmdefault}{\mddefault}{\updefault}Truth-membership}}}}}
\put(1062,1887){\makebox(0,0)[lb]{\smash{{{\SetFigFont{6}{7.2}{\rmdefault}{\mddefault}{\updefault}Function}}}}}
\put(1062,687){\makebox(0,0)[lb]{\smash{{{\SetFigFont{6}{7.2}{\rmdefault}{\mddefault}{\updefault}Falsity-}}}}}
\put(1062,537){\makebox(0,0)[lb]{\smash{{{\SetFigFont{6}{7.2}{\rmdefault}{\mddefault}{\updefault}membership}}}}}
\put(1062,387){\makebox(0,0)[lb]{\smash{{{\SetFigFont{6}{7.2}{\rmdefault}{\mddefault}{\updefault}Function}}}}}
\put(87,1437){\makebox(0,0)[lb]{\smash{{{\SetFigFont{6}{7.2}{\rmdefault}{\mddefault}{\updefault}Input}}}}}
\put(2262,1437){\makebox(0,0)[lb]{\smash{{{\SetFigFont{6}{7.2}{\rmdefault}{\mddefault}{\updefault}Neutrosophic }}}}}
\put(2262,1287){\makebox(0,0)[lb]{\smash{{{\SetFigFont{6}{7.2}{\rmdefault}{\mddefault}{\updefault}Inference}}}}}
\put(3387,1212){\makebox(0,0)[lb]{\smash{{{\SetFigFont{6}{7.2}{\rmdefault}{\mddefault}{\updefault}Reducer}}}}}
\put(3387,1362){\makebox(0,0)[lb]{\smash{{{\SetFigFont{6}{7.2}{\rmdefault}{\mddefault}{\updefault}Type}}}}}
\put(3312,1512){\makebox(0,0)[lb]{\smash{{{\SetFigFont{6}{7.2}{\rmdefault}{\mddefault}{\updefault}Neutrosophic }}}}}
\put(4287,1362){\makebox(0,0)[lb]{\smash{{{\SetFigFont{6}{7.2}{\rmdefault}{\mddefault}{\updefault}Denutrosophication}}}}}
\put(5187,1512){\makebox(0,0)[lb]{\smash{{{\SetFigFont{6}{7.2}{\rmdefault}{\mddefault}{\updefault}Crisp Output}}}}}
\end{picture}
}
\end{center}
\caption{General Scheme of an INLS}
\end{figure}
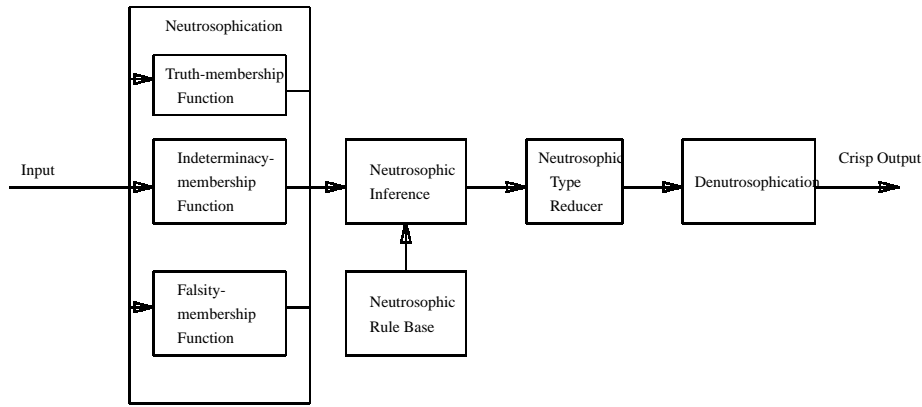

Suppose the neutrosophic rule base consists of $M$ rules in which each rule
has $n$ antecedents and one consequent. Let the $k$th rule be denoted 
by $R^k$ such that IF 
$x_1$ is $A_{1}^k$, $x_2$ is $A_{2}^k$, $\ldots$, and $x_n$ is $A_{n}^k$, THEN $y$
is $B^k$. 
$A_{i}^k$ is an interval neutrosophic set defined on universe $X_i$ with 
truth-membership
function $T_{A_{i}^k}(x_i)$, indeterminacy-membership function 
$I_{A_{i}^k}(x_i)$ and falsity-membership function $F_{A_{i}^k}(x_i)$, where
$T_{A_{i}^k}(x_i), I_{A_{i}^k}(x_i), F_{A_{i}^k}(x_i) \subseteq [0, 1], 1 \leq i \leq n$. 
$B^k$ is
an interval neutrosophic set defined on universe $Y$ with truth-membership
function $T_{B^k}(y)$,
indeterminacy-membership function $I_{B^k}(y)$ and falsity-membership function
$F_{B^k}(y)$, where $T_{B^k}(y), I_{B^k}(y), F_{B^k}(y) \subseteq [0, 1]$.
Given fact $x_1$ is $\tilde{A}_{1}^k, x_2$ is $\tilde{A}_{2}^k, \ldots
$,
and $x_n$ is $\tilde{A}_{n}^k$, then consequence $y$ is $\tilde{B}^k$.
$\tilde{A}_{i}^k$ is an interval neutrosophic set defined on universe
$X_i$ with 
truth-membership function $T_{\tilde{A}_{i}^k}(x_i)$, indeterminacy-membership
function $I_{\tilde{A}_{i}^k}(x_i)$ and falsity-membership function 
$F_{\tilde{A}_{i}^k}(x_i)$, where $T_{\tilde{A}_{i}^k}(x_i), I_{\tilde{A}_{i}^k}(x_i), F_{\tilde{A}_{i}^k}(x_i) \subseteq [0, 1], 1 \leq i \leq n$. 
$\tilde{B}^k$ is an interval neutrosophic set defined on universe $Y$ with 
truth-membership function
$T_{\tilde{B}^k}(y)$, indeterminacy-membership function $I_{\tilde{B}^k}(y)$
and falsity-membership function $F_{\tilde{B}^k}(y)$, where $T_{\tilde{B}^k}(y), I_{\tilde{B}^k}(y), F_{\tilde{B}^k}(y) \subseteq [0, 1]$. 
In this paper, we consider $a_i \leq X_i \leq b_i$ and $\alpha \leq Y \leq \beta$. 

An unconditional neutrosophic proposition is expressed by the phrase: 
``$Z$ is $C$", where $Z$ is a variable that receives values $z$ from a universal set $U$, and $C$ is an interval neutrosophic set defined on $U$ that represents a neutrosophic predicate. Each neutrosophic proposition $p$ is associated
with $\langle t(p), i(p), f(p) \rangle$ with $t(p), i(p), f(p) \subseteq [0, 1]$. In general, for any value $z$ of $Z$, $\langle t(p), i(p), f(p) \rangle = \langle T_{C}(z), I_{C}(z), F_{C}(z) \rangle$.
 		
For implication $p \rightarrow q$, we define the semantics as:
\begin{eqnarray}
\sup t_{p \rightarrow q} &=& \min (\sup t(p), \sup t(q)), \\
\inf t_{p \rightarrow q} &=& \min (\inf t(p), \inf t(q)), \\
\sup i_{p \rightarrow q} &=& \min (\sup i(p), \sup i(q)), \\
\inf i_{p \rightarrow q} &=& \min (\inf i(p), \inf i(q)), \\
\sup f_{p \rightarrow q} &=& \max (\sup f(p), \sup f(q)), \\
\inf f_{p \rightarrow q} &=& \max (\inf f(p), \inf f(q)),
\end{eqnarray}
where $t(p),i(p),f(p),t(q),i(q),f(q) \subseteq [0,1]$. 

Let $X = X_1 \times \cdots \times X_n$. 
The truth-membership function, indeterminacy-membership function and 
falsity-membership function $T_{\tilde{B}^k}(y), I_{\tilde{B}^k}(y), F_{\tilde{B}^k}(y)$ of a fired $k$th rule can be represented using the definition of 
interval neutrosophic composition functions (23--25) and the 
semantics of conjunction and disjunction defined in Table I and 
equations (38--43) as:
\begin{eqnarray}
\sup T_{\tilde{B}^k}(y) &=& \sup_{x \in X}\min(\sup T_{\tilde{A}_{1}^k}(x_1), \sup T_{A_{1}^k}(x_1), \ldots, \sup T_{\tilde{A}_{n}^k}(x_n), \sup T_{A_{n}^k}(x_n), \sup T_{B^k}(y)), \\ 
\inf T_{\tilde{B}^k}(y) &=& \sup_{x \in X}\min(\inf T_{\tilde{A}_{1}^k}(x_1), \inf T_{A_{1}^k}(x_1), \ldots, \inf T_{\tilde{A}_{n}^k}(x_n), \inf T_{A_{n}^k}(x_n), \inf T_{B^k}(y)), \\
\sup I_{\tilde{B}^k}(y) &=& \sup_{x \in X}\min(\sup I_{\tilde{A}_{1}^k}(x_1), \sup I_{A_{1}^k}(x_1), \ldots, \sup I_{\tilde{A}_{n}^k}(x_n), \sup I_{A_{n}^k}(x_n), \sup I_{B^k}(y)), \\ 
\inf I_{\tilde{B}^k}(y) &=& \sup_{x \in X}\min(\inf I_{\tilde{A}_{1}^k}(x_1), \inf I_{A_{1}^k}(x_1), \ldots, \inf I_{\tilde{A}_{n}^k}(x_n), \inf I_{A_{n}^k}(x_n), \inf I_{B^k}(y)), \\
\sup F_{\tilde{B}^k}(y) &=& \inf_{x \in X}\max(\sup F_{\tilde{A}_{1}^k}(x_1), \sup F_{A_{1}^k}(x_1), \ldots, \sup F_{\tilde{A}_{n}^k}(x_n), \sup F_{A_{n}^k}(x_n), \sup F_{B^k}(y)), \\
\inf F_{\tilde{B}^k}(y) &=& \inf_{x \in X}\max(\inf F_{\tilde{A}_{1}^k}(x_1), \inf F_{A_{1}^k}(x_1), \ldots, \inf F_{\tilde{A}_{n}^k}(x_n), \inf F_{A_{n}^k}(x_n), \inf F_{B^k}(y)),
\end{eqnarray}
where $y \in Y$.

Now, we give the algorithmic description of NLS. 

BEGIN

Step 1: Neutrosophication

Let $G^k$ be an interval neutrosophic set to represent the result of
the input and antecedent operation (neutrosophication) for $k$th rule, then
\begin{eqnarray}
\sup T_{G^k}(x) &=& \sup_{x \in X}\min(\sup T_{\tilde{A}_{1}^k}(x_1), \sup T_{A_{1}^k}(x_1), \ldots, \sup T_{\tilde{A}_{n}^k}(x_n), \sup T_{A_{n}^k}(x_n)), \\ 
\inf T_{G^k}(x) &=& \sup_{x \in X}\min(\inf T_{\tilde{A}_{1}^k}(x_1), \inf T_{A_{1}^k}(x_1), \ldots, \inf T_{\tilde{A}_{n}^k}(x_n), \inf T_{A_{n}^k}(x_n)), \\ 
\sup I_{G^k}(x) &=& \sup_{x \in X}\min(\sup I_{\tilde{A}_{1}^k}(x_1), \sup I_{A_{1}^k}(x_1), \ldots, \sup I_{\tilde{A}_{n}^k}(x_n), \sup I_{A_{n}^k}(x_n)), \\
\inf I_{G^k}(x) &=& \sup_{x \in X}\min(\inf I_{\tilde{A}_{1}^k}(x_1), \inf I_{A_{1}^k}(x_1), \ldots, \inf I_{\tilde{A}_{n}^k}(x_n), \inf I_{A_{n}^k}(x_n)), \\
\sup F_{G^k}(x) &=& \inf_{x \in X}\max(\sup F_{\tilde{A}_{1}^k}(x_1), \sup F_{A_{1}^k}(x_1), \ldots, \sup F_{\tilde{A}_{n}^k}(x_n), \sup F_{A_{n}^k}(x_n)), \\
\inf F_{G^k}(x) &=& \inf_{x \in X}\max(\inf F_{\tilde{A}_{1}^k}(x_1), \inf F_{A_{1}^k}(x_1), \ldots, \inf F_{\tilde{A}_{n}^k}(x_n), \inf F_{A_{n}^k}(x_n)),
\end{eqnarray}
where $x_i \in X_i$.

If $x_1, \ldots, x_n$ are crisp inputs, then equations (56--61) are simplified 
to
\begin{eqnarray}
\sup T_{G^k}(x) &=& \min(
\sup T_{A_{1}^k}(x_1), \ldots, \sup T_{A_{n}^k}
(x_n)), \\
\inf T_{G^k}(x) &=& \min( 
\inf T_{A_{1}^k}(x_1), \ldots, \inf T_{A_{n}^k}
(x_n)), \\
\sup I_{G^k}(x) &=& \min(
\sup I_{A_{1}^k}(x_1), \ldots, \sup I_{A_{n}^k}
(x_n)), \\
\inf I_{G^k}(x) &=& \min(
\inf I_{A_{1}^k}(x_1), \ldots, \inf I_{A_{n}^k}
(x_n)), \\
\sup F_{G^k}(x) &=& \max(
\sup F_{A_{1}^k}(x_1), \ldots, \sup F_{A_{n}^k}
(x_n)), \\
\inf F_{G^k}(x) &=& \max(
\inf F_{A_{1}^k}(x_1), \ldots, \inf F_{A_{n}^k}
(x_n)),
\end{eqnarray}
where $x_i \in X_i$.

Fig. 2 shows the conceptual diagram for neutrosophication of crisp inputs. 

\begin{figure}[htbp]
\label{figure2}
\begin{center}
\setlength{\unitlength}{0.00083333in}
\begingroup\makeatletter\ifx\SetFigFont\undefined%
\gdef\SetFigFont#1#2#3#4#5{%
  \reset@font\fontsize{#1}{#2pt}%
  \fontfamily{#3}\fontseries{#4}\fontshape{#5}%
  \selectfont}%
\fi\endgroup%
{\renewcommand{\dashlinestretch}{30}
\begin{picture}(2911,5388)(0,-10)
\path(675,4011)(2250,4011)
\blacken\path(2130.000,3981.000)(2250.000,4011.000)(2130.000,4041.000)(2130.000,3981.000)
\path(675,4011)(675,5361)
\blacken\path(705.000,5241.000)(675.000,5361.000)(645.000,5241.000)(705.000,5241.000)
\path(675,2286)(675,3636)
\blacken\path(705.000,3516.000)(675.000,3636.000)(645.000,3516.000)(705.000,3516.000)
\path(675,2286)(2250,2286)
\blacken\path(2130.000,2256.000)(2250.000,2286.000)(2130.000,2316.000)(2130.000,2256.000)
\path(675,486)(675,1836)
\blacken\path(705.000,1716.000)(675.000,1836.000)(645.000,1716.000)(705.000,1716.000)
\path(675,486)(2250,486)
\blacken\path(2130.000,456.000)(2250.000,486.000)(2130.000,516.000)(2130.000,456.000)
\path(675,4011)(1200,4911)(1800,4011)
\path(975,4011)(1200,4911)(1500,4011)
\path(675,1236)(1050,486)(1800,1161)
\path(675,1236)(1125,786)(1800,1161)
\path(675,2286)(1050,2811)(1350,2286)
	(1725,2886)(1950,2286)
\path(675,2286)(1050,2586)(1350,2286)
	(1725,2661)(1950,2286)
\dashline{60.000}(675,4911)(2025,4911)
\dashline{60.000}(675,3186)(2100,3186)
\dashline{60.000}(675,1461)(2100,1461)
\path(900,486)(900,4836)
\path(900,4386)(2775,4386)
\path(900,2586)(2775,2586)
\path(900,2436)(2775,2436)
\path(900,1011)(2700,1011)
\path(900,786)(2700,786)
\put(150,5286){\makebox(0,0)[lb]{\smash{{{\SetFigFont{6}{7.2}{\rmdefault}{\mddefault}{\updefault}Truth-}}}}}
\put(150,5136){\makebox(0,0)[lb]{\smash{{{\SetFigFont{6}{7.2}{\rmdefault}{\mddefault}{\updefault}membership}}}}}
\put(150,4986){\makebox(0,0)[lb]{\smash{{{\SetFigFont{6}{7.2}{\rmdefault}{\mddefault}{\updefault}Function}}}}}
\put(0,3561){\makebox(0,0)[lb]{\smash{{{\SetFigFont{6}{7.2}{\rmdefault}{\mddefault}{\updefault}Indeterminacy-}}}}}
\put(0,3411){\makebox(0,0)[lb]{\smash{{{\SetFigFont{6}{7.2}{\rmdefault}{\mddefault}{\updefault}membership}}}}}
\put(0,3261){\makebox(0,0)[lb]{\smash{{{\SetFigFont{6}{7.2}{\rmdefault}{\mddefault}{\updefault}Function}}}}}
\put(600,3936){\makebox(0,0)[lb]{\smash{{{\SetFigFont{6}{7.2}{\rmdefault}{\mddefault}{\updefault}0}}}}}
\put(600,4911){\makebox(0,0)[lb]{\smash{{{\SetFigFont{6}{7.2}{\rmdefault}{\mddefault}{\updefault}1}}}}}
\put(600,2211){\makebox(0,0)[lb]{\smash{{{\SetFigFont{6}{7.2}{\rmdefault}{\mddefault}{\updefault}0}}}}}
\put(600,3111){\makebox(0,0)[lb]{\smash{{{\SetFigFont{6}{7.2}{\rmdefault}{\mddefault}{\updefault}1}}}}}
\put(600,411){\makebox(0,0)[lb]{\smash{{{\SetFigFont{6}{7.2}{\rmdefault}{\mddefault}{\updefault}0}}}}}
\put(600,1461){\makebox(0,0)[lb]{\smash{{{\SetFigFont{6}{7.2}{\rmdefault}{\mddefault}{\updefault}1}}}}}
\put(2325,3936){\makebox(0,0)[lb]{\smash{{{\SetFigFont{6}{7.2}{\rmdefault}{\mddefault}{\updefault}X}}}}}
\put(2325,2211){\makebox(0,0)[lb]{\smash{{{\SetFigFont{6}{7.2}{\rmdefault}{\mddefault}{\updefault}X}}}}}
\put(2325,411){\makebox(0,0)[lb]{\smash{{{\SetFigFont{6}{7.2}{\rmdefault}{\mddefault}{\updefault}X}}}}}
\put(900,336){\makebox(0,0)[lb]{\smash{{{\SetFigFont{6}{7.2}{\rmdefault}{\mddefault}{\updefault}x}}}}}
\put(2850,4161){\makebox(0,0)[lb]{\smash{{{\SetFigFont{6}{7.2}{\rmdefault}{\mddefault}{\updefault}T}}}}}
\put(2850,2436){\makebox(0,0)[lb]{\smash{{{\SetFigFont{6}{7.2}{\rmdefault}{\mddefault}{\updefault}I}}}}}
\put(2775,861){\makebox(0,0)[lb]{\smash{{{\SetFigFont{6}{7.2}{\rmdefault}{\mddefault}{\updefault}F}}}}}
\put(1125,36){\makebox(0,0)[lb]{\smash{{{\SetFigFont{8}{9.6}{\rmdefault}{\mddefault}{\updefault}Neutrosophication}}}}}
\put(75,1686){\makebox(0,0)[lb]{\smash{{{\SetFigFont{6}{7.2}{\rmdefault}{\mddefault}{\updefault}membership}}}}}
\put(75,1536){\makebox(0,0)[lb]{\smash{{{\SetFigFont{6}{7.2}{\rmdefault}{\mddefault}{\updefault}Function}}}}}
\put(75,1836){\makebox(0,0)[lb]{\smash{{{\SetFigFont{6}{7.2}{\rmdefault}{\mddefault}{\updefault}Falsity}}}}}
\end{picture}
}
\end{center}
\caption{Conceputal Diagram for Neutrosophication of Crisp Inputs}
\end{figure}
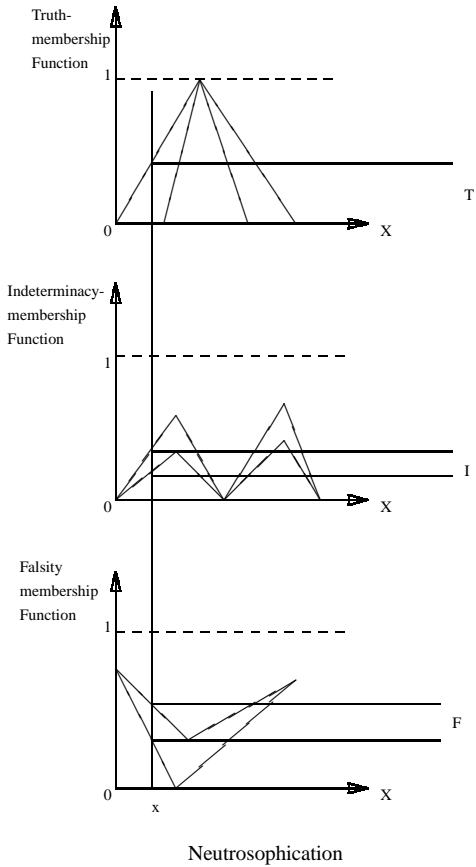

Step 2: Neutrosophic Inference

The core of NLS is the neutrosophic inference, the principle of which has
already been explained above. Suppose the $k$th rule is fired. Here we 
restate the result: 

\begin{eqnarray}
\sup T_{\tilde{B}^k}(y) &=& 
\min(\sup T_{G^k}(x), \sup T_{B^k}(y)), \\
\inf T_{\tilde{B}^k}(y) &=& 
\min(\inf T_{G^k}(x), \inf T_{B^k}(y)), \\
\sup I_{\tilde{B}^k}(y) &=& 
\min(\sup I_{G^k}(x), \sup I_{B^k}(y)), \\
\inf I_{\tilde{B}^k}(y) &=& 
\min(\inf I_{G^k}(x), \inf I_{B^k}(y)), \\
\sup F_{\tilde{B}^k}(y) &=& 
\max(\sup F_{G^k}(x), \sup F_{B^k}(y)), \\
\inf F_{\tilde{B}^k}(y) &=& 
\max(\inf F_{G^k}(x), \inf F_{B^k}(y)),
\end{eqnarray}
where $x \in X, y \in Y$.

Suppose that $N$ rules in the neutrosophic rule base are fired, where $N \leq M$
; then,
the output interval neutrosophic set $\tilde{B}$ is:
\begin{eqnarray}
\sup T_{\tilde{B}}(y) &=& \max_{k=1}^{N} \sup T_{\tilde{B}^k}(y), \\
\inf T_{\tilde{B}}(y) &=& \max_{k=1}^{N} \inf T_{\tilde{B}^k}(y), \\
\sup I_{\tilde{B}}(y) &=& \max_{k=1}^{N} \sup I_{\tilde{B}^k}(y), \\
\inf I_{\tilde{B}}(y) &=& \max_{k=1}^{N} \inf I_{\tilde{B}^k}(y), \\
\sup F_{\tilde{B}}(y) &=& \min_{k=1}^{N} \sup T_{\tilde{B}^k}(y), \\
\inf T_{\tilde{B}}(y) &=& \min_{k=1}^{N} \inf T_{\tilde{B}^k}(y),
\end{eqnarray}
where $y \in Y$.

Step 3: Neutrosophic type reduction

After neutrosophic inference, we will get an interval neutrosophic set $\tilde{B}$ with $T_{\tilde{B}}(y), I_{\tilde{B}}(y), F_{\tilde{B}}(y) \subseteq [0, 1]$. Then, we do the neutrosophic type reduction to transform each interval 
into one number. There are many ways to do it, here, we give one method:
\begin{eqnarray}
T_{\tilde{B}}^{'}(y) &=& (\inf T_{\tilde{B}}(y) + \sup T_{\tilde{B}}(y)) / 2, \\
I_{\tilde{B}}^{'}(y) &=& (\inf I_{\tilde{B}}(y) + \sup I_{\tilde{B}}(y)) / 2, \\
F_{\tilde{B}}^{'}(y) &=& (\inf F_{\tilde{B}}(y) + \sup F_{\tilde{B}}(y)) / 2,
\end{eqnarray} 
where $y \in Y$.

So, after neutrosophic type reduction, we will get an ordinary neutrosophic set
(a type-1 neutrosophic set)
$\tilde{B}$. Then we need to do the deneutrosophication to get a crisp output.

Step 4: Deneutrosophication

The purpose of 
deneutrosophication is to convert an ordinary neutrosophic set (a type-1 neutrosophic set) obtained by
neutrosophic type reduction to a single real number which represents the
real output. 
Similar to defuzzification~\cite{KY95}, there are 
many deneutrosophication methods according to different applications. Here
we give one method. The deneutrosophication process consists of two steps. 

Step 4.1: \emph{Synthesization}: It is the process to transform an ordinary neutrosophic set (a type-1 neutrosophic set) $\tilde{B}$ into a fuzzy set $\bar{B}$.
It can be expressed using the following function: 

\begin{equation}
f(T_{\tilde{B}}^{'}(y), I_{\tilde{B}}^{'}(y), F_{\tilde{B}}^{'}(y)) : [0,1] \times [0,1] \times [0,1] \rightarrow [0,1]
\end{equation}

Here we give one definition of $f$:

\begin{equation}
T_{\bar{B}}(y) = a*T_{\tilde{B}}^{'}(y) + b*(1-F_{\tilde{B}}^{'}(y)) + c*I_{\tilde{B}}^{'}(y)/2+d*(1-I_{\tilde{B}}^{'}(y)/2),
\end{equation}
where $0 \leq a,b,c,d \leq 1, a+b+c+d = 1$.

The purpose of synthesization is to calculate the overall truth degree according to three components: truth-membership function, indeterminacy-membership function and falsity-membership function. The component--truth-membership function
gives the direct information about the truth-degree, so we use it directly in
the formula; The component--falsity-membership function gives the indirect
information about the truth-degree, so we use $(1 - F)$ in the formula. To
understand the meaning of indeterminacy-membership function $I$, 
we give an example: a statement is ``The quality of service is good", now firstly a person has to select a decision among $\{T, I, F\}$, 
secondly he or she has to answer the degree of the decision in $[0, 1]$. 
If he or she chooses $I = 1$, it means $100\%$ ``not sure" about the statement, 
i.e., $50\%$ true and $50\%$ false for the statement ($100\%$ balanced), 
in this 
sense, $I = 1$ contains the potential truth value $0.5$. If he or she chooses
$I = 0$, it means $100\%$ ``sure" about the statement, i.e., either $100\%$
true or $100\%$ false for the statement ($0\%$ balanced), in this sense, 
$I = 0$
is related to two extreme cases, but we do not know which one is in his or 
her mind.
So we have to consider both at the same time: $I = 0$ contains the potential
truth value that is either $0$ or $1$. If $I$ decreases from $1$ to $0$, then 
the potential truth value changes from one value $0.5$ to two different possible values gradually to the final possible ones $0$ and 
$1$ (i.e., from $100\%$ balanced to $0\%$ balanced), 
since he or she does not choose either $T$ or $F$ but $I$, we do not know 
his or her
final truth value. Therefore, the formula has  to consider two potential 
truth 
values implicitly represented by $I$ with different weights ($c$ and $d$) 
because of lack of 
his or her final decision information after he or she has chosen $I$.
Generally, $a > b > c, d$; $c$ and $d$ could be decided subjectively or 
objectively as long as enough information is available.
The parameters $a,b,c$ and $d$ can be tuned using learning algorithms such as
neural networks and genetic algorithms in the 
development of 
application to
improve the performance of the NLS.

Step 4.2: \emph{Calculation of a typical neutrosophic value}: 
Here we introduce one method of calculation of center of area. The method is sometimes called 
the 
\emph{center of gravity method} or \emph{centroid method}, the 
deneutrosophicated value, 
$dn(T_{\bar{B}}(y))$ is 
calculated by the formula
\begin{equation}
  dn(T_{\bar{B}}(y)) = \frac{\int_{\alpha}^{\beta} T_{\bar{B}}(y)y dy}{\int_{\alpha}^{\beta} T_{\bar{B}} dy}. 
\end{equation} 

END.

\section{Conclusions}
\label{conclusion}
In this paper, we give the formal definition of interval neutrosophic logic
which are unifying framework of many other classical logics such as fuzzy
logic, intuitionistic fuzzy logic and paraconsistent logics, etc. 
Interval neutrosophic logic include
interval neutrosophic propositional logic and first order interval 
neutrosophic predicate, we call them classical (standard) neutrosophic logic.
In the future, we also will discuss and explore the non-classical (non-standard) neutrosophic logic such as modal interval neutrosophic logic, temporal interval neutrosophic logic, etc. Interval neutrosophic logic can not only handle
imprecise, fuzzy and imcomplete propositions but also inconsistent propositions without the danger of trivilization. The paper also give one application based
on the semantic notion of interval neutrosophic logic -- the interval neutrosophic logic systems (INLS) which
are the generalization of classical FLS and interval valued fuzzy FLS. 
Interval neutrosophic logic will have a lot of potential applications in 
computational Web intelligence~\cite{ZKL04}. For example, current fuzzy Web 
intelligence techniques can be improved by using more reliable interval neutrosophic logic methods because $T, I$ and $F$ are all used in decision making. In large, such robust interval neutrosophic logic methods can also be used in other applications such as medical informatics, bioinformatics and human-oriented decision-making under uncertainty. In fact, interval neutrosophic sets and interval neutrosophic
logic could be applied in the fields that fuzzy sets and fuzz logic are suitable for, also the fields that paraconsistent logics are suitable for.
 
\section*{Acknowledgment}
The authors would like to thank Mr. F. H. Jiang for his valuable suggestions. 



\bibliographystyle{IEEEtran}
\bibliography{IEEEfull,ref}

\begin{thebibliography}{10}
\providecommand{\url}[1]{#1}
\csname url@rmstyle\endcsname
\providecommand{\newblock}{\relax}
\providecommand{\bibinfo}[2]{#2}
\providecommand\BIBentrySTDinterwordspacing{\spaceskip=0pt\relax}
\providecommand\BIBentryALTinterwordstretchfactor{4}
\providecommand\BIBentryALTinterwordspacing{\spaceskip=\fontdimen2\font plus
\BIBentryALTinterwordstretchfactor\fontdimen3\font minus
  \fontdimen4\font\relax}
\providecommand\BIBforeignlanguage[2]{{%
\expandafter\ifx\csname l@#1\endcsname\relax
\typeout{** WARNING: IEEEtran.bst: No hyphenation pattern has been}%
\typeout{** loaded for the language `#1'. Using the pattern for}%
\typeout{** the default language instead.}%
\else
\language=\csname l@#1\endcsname
\fi
#2}}

\bibitem{ZAD65}
L.~Zadeh, ``Fuzzy sets,'' \emph{Inform. and Control}, vol.~8, pp. 338--353,
  1965.

\bibitem{TUR86}
I.~Turksen, ``Interval valued fuzzy sets based on normal forms,'' \emph{Fuzzy
  Sets and Systems}, vol.~20, pp. 191--210, 1986.

\bibitem{KM98}
N.~Karnik and J.~Mendel, ``Introduction to type-2 fuzzy logic systems,'' in
  \emph{Proc. 1998 IEEE Fuzz Conf.}, 1998, pp. 915--920.

\bibitem{LM00}
Q.~Liang and J.~Mendel, ``Interval type-2 fuzzy logic systems: theory and
  design,'' \emph{IEEE Transactions On Fuzzy Systems}, vol.~8, pp. 535--550,
  2000.

\bibitem{MJ02}
J.~Mendel and R.~John, ``Type-2 fuzzy sets made simple,'' \emph{IEEE
  Transactions on Fuzzy Systems}, vol.~10, pp. 117--127, 2002.

\bibitem{ATA86}
K.~Atanassov, ``Intuitionistic fuzzy sets,'' \emph{Fuzzy Sets and Systems},
  vol.~20, pp. 87--96, 1986.

\bibitem{ATA88}
------, ``Two variants of intuitionistic fuzzy propositional calculus,'' 1988,
  preprint IM-MFAIS-5-88.

\bibitem{ATA90}
K.~Atanassov and G.~Gargov, ``Intuitionistic fuzzy logic,'' \emph{Compt. Rend.
  Acad. Bulg. Sci.}, vol.~43, pp. 9--12, 1990.

\bibitem{ATA98}
------, ``Elements of intuitionistic fuzzy logic. part i,'' \emph{Fuzzy Sets
  and Systems}, vol.~95, pp. 39--52, 1998.

\bibitem{ACM02}
S.~de~Amo, W.~Carnielli, and J.~Marcos, ``A logical framework for integrating
  inconsistent information in multiple databases,'' in \emph{Proc. PoIKS'02,
  LNCS 2284}, 2002, pp. 67--84.

\bibitem{COS77}
N.~Costa, ``On the theory of inconsistent formal systems,'' \emph{Notre Dame
  Journal of Formal Logic}, vol.~15, pp. 621--630, 1977.

\bibitem{BEL77}
N.~D. Belnap, ``A useful four-valued logic,'' in \emph{Modern Uses of
  Many-valued Logic}, G.~Eppstein and J.~M. Dunn, Eds.\hskip 1em plus 0.5em
  minus 0.4em\relax Reidel, Dordrecht, 1977, pp. 8--37.

\bibitem{SMA03}
F.~Smarandache, \emph{A Unifying Field in Logics: Neutrosophic Logic.
  Neutrosophy, Neutrosophic Set, Neutrosophic Probability and
  Statistics}.\hskip 1em plus 0.5em minus 0.4em\relax Phoenix: Xiquan, 2003,
  third edition.

\bibitem{WPZR04}
\BIBentryALTinterwordspacing
H.~Wang, P.~Madiraju, Y.~Zhang, and R.~Sunderraman, ``Interval neutrosophic
  sets,'' 2004. [Online]. Available: \url{xxx.lanl.gov/pdf/math.GM/0409113}
\BIBentrySTDinterwordspacing

\bibitem{SMA01}
F.~Smarandache, ``Definitions derived from neutrosophics,'' in \emph{First
  International Conference on Neutrosophy, Neutrosophic Logic, Neutrosophic
  Set, Neutrosophic Probability and Statistics}, University of New Mexico,
  Gallup, December 2001.

\bibitem{WZR04}
H.~B. Wang, Y.~Q. Zhang, and R.~Sunderraman, ``Soft semantic web services
  agent,'' in \emph{The Proceedings of NAFIPS 2004}, 2004, pp. 126--129.

\bibitem{MEN87}
E.~Mendelson, \emph{Introduction to Mathematical Logic}.\hskip 1em plus 0.5em
  minus 0.4em\relax Princeton, NJ: Van Nostrand, 1987, third edition.

\bibitem{KY95}
G.~J. Klir and B.~Yuan, \emph{Fuzzy Sets and Fuzzy Logic: Theory and
  Applications}, Upper Saddle River, New Jersey, 1995.

\bibitem{ZKL04}
Y.-Q. Zhang, A.~Kandel, T.~Lin, and Y.~Yao, \emph{Computational Web
  Intelligence: Intelligent Technology for Web Applications, Series in Machine
  Perception and Artificial Intelligence}.\hskip 1em plus 0.5em minus
  0.4em\relax World Scientific, 2004, volume 58.

\end{thebibliography}









\end{document}